*universe*

MDPI

Article

# The Origin of Intergalactic Light in Compact Groups of Galaxies

Mark J. Henriksen

University of Maryland, Baltimore County, Physics Department, Baltimore MD, 21250; henrikse@umbc.edu

**Abstract:** We investigate the origin of intergalactic light (IGL) in close groups of galaxies. IGL is hypothesized to be the byproduct of interaction and merger within compact groups. Comparing the X-ray point source population in our sample of compact groups that have intergalactic light with compact groups without IGL, we find marginal evidence for a small increase in ultra-luminous X-ray sources (ULXs). There is also a significant bias towards lower luminosity high mass X-ray binaries (HMXRBs). We interpret this as an indication that groups with visible IGL represent a later evolutionary phase than other compact groups. They have galaxies characterized by quenching of star formation (lower star formation rate (SFR) inferred from lower HMXRB luminosity) after stellar material has been removed from the galaxies into the intergalactic medium, which is the source of the IGL. We conclude that the presence of an increased fraction of ULXs is due to past interaction and mergers within groups that have IGL.

## 1. Introduction

Compact Groups (CGs) of Galaxies are small and relatively isolated aggregates with mean separations on the order of a component galaxy diameter. They are the densest galaxy concentrations found in some of the lowest density regions of large scale structure. CGs represent an important link in our understanding of galaxy evolution, specifically evolution at the highest spatial frequency scales of the non-clustered structure because these often strange configurations (e.g., the VV 172 group shown in Figure 1) sometimes show signs of interaction, such as an optical envelope [1]. They are excellent laboratories for studying extreme effects of interaction (e.g., stripping, tidal interaction, star formation/AGN stimulation, secular evolution, shocks) and merger thought to be important at earlier epochs. Well studied local analogs are therefore fundamentally important for interpreting interaction related phenomena at high redshift, especially in the context of the origin of massive ellipticals.

Giant ellipticals with high X-ray luminosity are hypothesized to be the endpoint of galaxy mergers that occur primarily within groups. Simulations [2] show that members of a galaxy group can merge to form a large elliptical galaxy in less than a Hubble time. The group X-ray emission would then be retained as a bright halo around a giant elliptical galaxy. The discovery of such an endpoint [3] was dubbed a "fossil group". Since then, a number of fossil groups have been found [4]. The qualifying feature of a fossil group is that it shows an overly X-ray luminous giant elliptical surrounded by substantially less optically bright group members. It is estimated that perhaps 10–20% of the groups fit this definition [5,6].

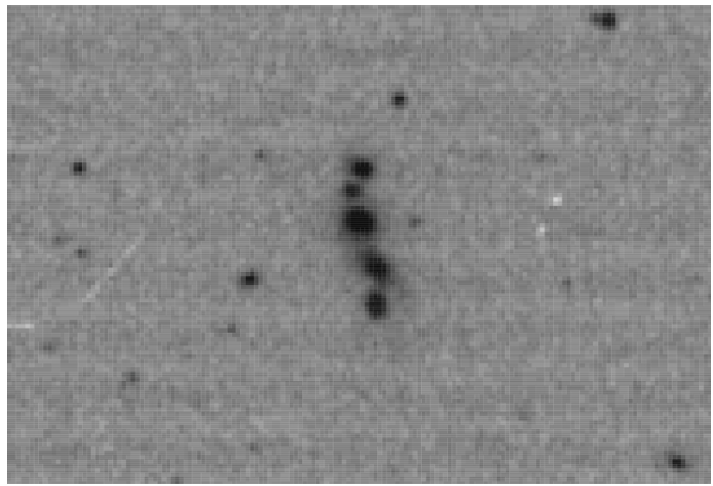

**Figure 1.** The highly unusual dynamical configuration and small size of HCG 55 appears to be a pre-merger stage.

In this paper, we describe a search for evidence that supports this scenario for fossil group formation by looking for an intermediate stage between a typical compact group and a merged remnant. Fossil groups were suggested to be the end products of the merging of L* galaxies in a low-density environment [5]. Thus we have chosen to study compact groups; groups of galaxies whose environment is relatively sparse and isolated. We propose that the "smoking gun" of the merged stage are the products of interaction and merger among group members. The IGL, seen in a number of compact groups, is likely a product of interaction and merger. This optical halo is hypothesized to be galactic material expelled in galactic winds driven by star formation within interacting galaxies or tidally stripped interstellar material.

The luminosity function of HMXRBs is strongly linked to recent star formation with $(L_x/4.0 \times 10^{39}\ erg\ s^{-1}) \sim SFR\ (M_\odot yr^{-1})$ [7]. Galaxies with high star formation rates were found to be at the high end of the HMXRB luminosity function (LF) and have luminosities of $10^{39}$–$10^{40}$ erg s$^{-1}$. By establishing the X-ray LF for point sources in compact groups with IGL, and thus inferring the SFR, we can test if IGL marks an important stage in the formation of a giant elliptical galaxy. Without evidence of interaction and merger in the compact group galaxy population, there is no link between the fossil and the group.

The best-known sample of CGs involves a visually selected catalog that contains 100 of the brightest low redshift CGs [8], abbreviated HCG. They typically involve 4–5 galaxies with mean separation r ~40 kpc and inferred 3D velocity dispersion ~300 km s$^{-1}$. While some CGs are chance projections, statistical studies involving, among other things, an HI deficit [9] and detailed studies of individual groups [10] make it clear that the majority are physically dense systems. A linear relation between velocity dispersion and X-ray luminosity extends over 6 decades of luminosity, ranging from clusters to groups, supporting that groups, such as clusters, are dynamically self-gravitating systems. Most relevant to this work is the low end of the X-ray luminosity < $10^{41}$ erg s$^{-1}$, where the correlation has a larger dispersion and therefore likely contains another component which is non-gravitational (e.g., stellar or AGN sources). Disadvantages of the Hickson catalog involve its incompleteness and selection biases associated with its compilation. In the past few years more complete catalogs have been compiled using digital techniques [11]. These catalogs both amplify the population of known bright CGs and make it possible to better interpret results obtained from the less complete Hickson catalog. The population of CGs within z = 0.1 has now increased to several hundred. The sample used in this paper is obtained from both catalogs, though is limited by a lack of spectroscopic redshifts for the non-HCGs.

## 2. Data and Methods

We test the evolutionary hypothesis by assembling a sample of compact groups with IGL. We characterize the SFR in these compact groups by their X-ray point source population. In low luminosity, early-type galaxies and spirals, the X-ray emission from X-ray Binaries (XRBs) dominates the diffuse X-ray component. Our sample, abbreviated IGL, has significant Galactic X-ray emission observed with modern observatories: Chandra and XMM. The spatial resolution of Chandra and XMM is needed to resolve and separate the galaxy point source components as the galaxies are often very close. By comparing groups with IGL with various other samples, we hope to gain insight on how the IGL relates to interaction and merger, and therefore the formation of giant ellipticals.

The distances of the groups range from 25–179 Mpc. At these distances, relatively few XRBs are detected in any single galaxy. This appears to be the case in the compilation of X-ray LFs (XLFs) in [12], though we note that XLFs have been published with as few as three sources. Therefore, we construct the XRB distribution from all of the galaxies in the groups, without distinguishing between groups or galaxy type, in order to acquire a representative sample. This is justified because we are comparing the XRB distribution in CGs with optical halos versus CGs without halos, and there is no galaxy-type bias in between samples.

We have constructed two samples of groups with intergalactic light. The first involves southern groups that meet the following criteria: (1) groups with at least 4 members, (2) galaxy separations on the order of 1-2 galactic diameters (40–80kpc), (3) accordant redshifts $\Delta V < 1000$ km s$^{-1}$ and (4) visual inspection of optical images of the members, HCG 6 and HCG 16 [13], HCG 40 [14], SCG0121-3521 and SCG2345-2824 [11] shows diffuse intergalactic light.

The second sample consists of HCGs reported to have IGL in the literature [15–17].

The sample luminosities cover the range, $10^{39}$–$10^{40}$ erg s$^{-1}$, typical of ULXs and stretches down a factor of 100 lower in luminosity to probe LMXBs and luminous AGN. A $L_X$ vs. $L_B$ correlation exists for gravitationally bound objects ranging from compact groups up to rich galaxy clusters. The low luminosity end, which is inhabited by compact groups with a high spiral fraction and weak X-ray, typically has a large spread in $L_x$. This indicates a non-gravitational component, which we investigate here.

The X-ray point source data given in Tables 1 and 2 are obtained using the online HEASARC tool Xassist. Xassist uses pipeline analysis streams for detecting and characterizing X-ray sources in Chandra and XMM data [18]. Chandra data is preferred and used when available due to its arcsec spatial resolution. When not available, XMM is used, which has several arc second spatial resolution, still adequate for this study. Xassist results are well documented and a complete, detailed explanation of the results and the automated analysis is given at the Xassist website. The count rate, flux and luminosity are in the 0.3–8 keV energy range. Count rate is converted to flux using a power-law model with photon index 1.8. The Galactic column density ($n_H$) that absorbs the model utilizes $n_H$ from the HEASoft tool. Distances and redshifts are obtained from the website: ned.Ipac.Caltech.edu

Table 1. Compact Groups with a specific selection criterion.

| Group–Source | Distance (Mpc) | Z | Counts | Error | Flux 10$^{-14}$ erg cm$^{-2}$ s$^{-1}$ | Lx (10$^{40}$) ergs s$^{-1}$ | Location |
|---|---|---|---|---|---|---|---|
| HCG 16-1 | 54.66 | 0.0132 | 152.94 | 13.38 | 7.80 | 2.71 | Core |
| HCG 16-2 | - | - | 69.95 | 16.89 | 3.73 | 1.29 | Disk |
| HCG 16-3 | - | - | 99.25 | 15.61 | 4.98 | 1.73 | Disk |
| HCG 16-4 | - | - | 242.74 | 22.82 | 12.0 | 4.16 | Core |
| HCG 16-5 | - | - | 229.45 | 15.72 | 11.6 | 4.02 | Core |
| SCG0121-3521-1 | | | 28.48 | 6.00 | 3.87 | - | |
| SCG0121-3521-2 | | | 12.67 | 3.56 | 1.29 | - | |
| SCG0121-3521-3 | | | 3.95 | 2.00 | 0.42 | - | |
| SCG0121-3521-4 | | | 31.65 | 6.34 | 4.06 | - | |
| SCG0121-3521-4 | | | 50.87 | 8.08 | 3.51 | - | |
| SCG0121-3521-4 | | | 627.74 | 25.79 | 38.5 | - | |
| SCG0121-3521-4 | | | 104.05 | 10.20 | 2630. | - | |
| SCG2345-2824-1 | | | 139.65 | 14.28 | 4.18 | - | |
| SCG2345-2824-1 | | | 144.08 | 17.52 | 4.80 | - | |
| SCG2345-2824-2 | | | 163.89 | 21.83 | 5.39 | - | |
| SCG2345-2824-3 | | | 31.81 | 6.48 | 1.11 | - | |
| HCG 40-1 | 103.63 | 0.0223 | 17.98 | 3.49 | 0.56 | 0.713 | Core |
| HCG 40-2 | - | - | 4.29 | 2.24 | 0.20 | 0.253 | Disk |
| HCG 40-3 | - | - | 7.00 | 3.00 | 0.15 | 0.188 | Core |
| HCG 40-4 | - | - | 6.21 | 3.22 | 0.34 | 0.429 | Core |
| HCG 40-5 | - | - | 82.92 | 5.91 | 2.20 | 2.79 | Core |
| HCG 6-1 | 162.69 | 0.0379 | 0.0034 | 0.0004 | 0.53 | 1.70 | Disk |
| HCG 6-2 | - | - | 0.028 | 0.0015 | 2.66 | 8.60 | Disk |

**Table 2.** HCGs with IGL reported in the literature.

| Group | Distance (Mpc) | Z | Counts | Error | Flux $10^{-14}$ erg cm$^{-2}$ s$^{-1}$ | $L_x$(x$10^{40}$) ergs s$^{-1}$ | Location |
|---|---|---|---|---|---|---|---|
| HCG 15-1 | 97 | 0.0228 | 80.88 | 14.42 | 1.50 | 1.67 | Disk |
| HCG 15-2 | - | - | 1108.82 | 49.98 | 35.0 | 38.86 | Core |
| HCG 35 | - | - | - | - | - | - | |
| HCG 44-1 | 25 | 0.0046 | 88.98 | 7.27 | 4.56 | 0.33 | Core |
| HCG 44-2 | - | - | 54.30 | 3.26 | 3.14 | 0.22 | Disk |
| HCG 44-3 | - | - | 62.28 | 8.45 | 11.3 | 0.81 | Core |
| HCG 51-1 | 118 | 0.0258 | 242.94 | 12.82 | 3.26 | 5.40 | Core |
| HCG 51-2 | - | - | 9.38 | 5.89 | 0.51 | 0.85 | Disk |
| HCG 51-3 | - | - | 34.33 | 10.27 | 7.12 | 11.80 | Disk |
| HCG 51-4 | - | - | 44.65 | 5.37 | 1.30 | 2.15 | Core |
| HCG 79-1 | 65 | 0.0145 | 64.1 | 8.8 | 0.31 | 0.225 | Core |
| HCG 79-2 | - | - | 12.91 | 5.09 | 0.38 | 0.273 | Disk |
| HCG 79-3 | - | - | 9.6 | 3.3 | 0.69 | 0.503 | Core |
| HCG 79-4 | - | - | 12.17 | 5.32 | 0.18 | 0.078 | Disk |
| HCG 88 | - | - | - | - | - | - | |
| HCG 90-1 | 34.87 | 0.0088 | 45362.5 | 216.22 | 2960. | 414.25 | Core |
| HCG 90-2 | - | - | 58.26 | 8.35 | 0.20 | 0.029 | Disk |
| HCG 90-3 | - | - | 92.59 | 12.47 | 1.21 | 0.169 | Disk |
| HCG 90-4 | - | - | 23.38 | 5.20 | 0.19 | 0.027 | Core |
| HCG 90-5 | - | - | 13.61 | 3.69 | 0.18 | 0.025 | Disk |
| HCG 90-6 | - | - | 19.39 | 5.02 | 0.86 | 0.120 | Core |
| HCG 90-7 | - | - | 7.46 | 2.83 | 0.21 | 0.030 | Disk |
| HCG 90-8 | - | - | 5.68 | 2.45 | 5.21 | 0.729 | Disk |
| HCG 90-9 | - | - | 6.29 | 2.65 | 4.88 | 0.683 | Core |
| HCG 94-1 | 179.03 | 0.0417 | 26.36 | 5.62 | 3.85 | 15.14 | Core |
| HCG 94-2 | - | - | 49.26 | 8.69 | 6.88 | 27.05 | Core |
| HCG 95-1 | 169.64 | 0.0396 | 0.0033 | 0.0004 | 0.32 | 1.16 | Core |
| HCG 95-2 | - | - | 0.0056 | 0.0005 | 1.55 | 5.43 | Core |
| HCG 95-3 | - | - | 0.029 | 0.0014 | 3.41 | 11.96 | Core |

## 3. Results

Figures 2–7 show the location of the X-ray sources on the Digitized Sky-Survey image of each compact group. The number of the X-ray source corresponds to the number in the Tables. By visual inspection, a high degree of interaction between several of the galaxies within the groups is apparent.

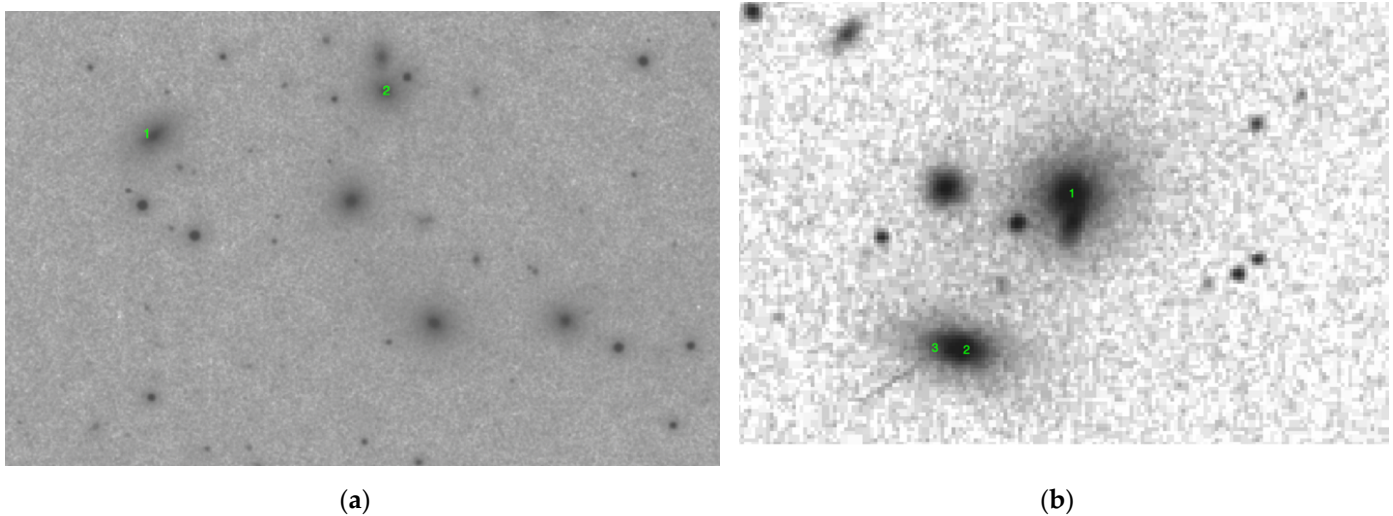

(**a**)      (**b**)

**Figure 2.** (**a**) XMM X-ray sources coincident with the galaxies on the DSS image of HCG 15. (**b**) Chandra X-ray point sources on selected galaxies in HCG 51 (missing group galaxies do not have point sources).

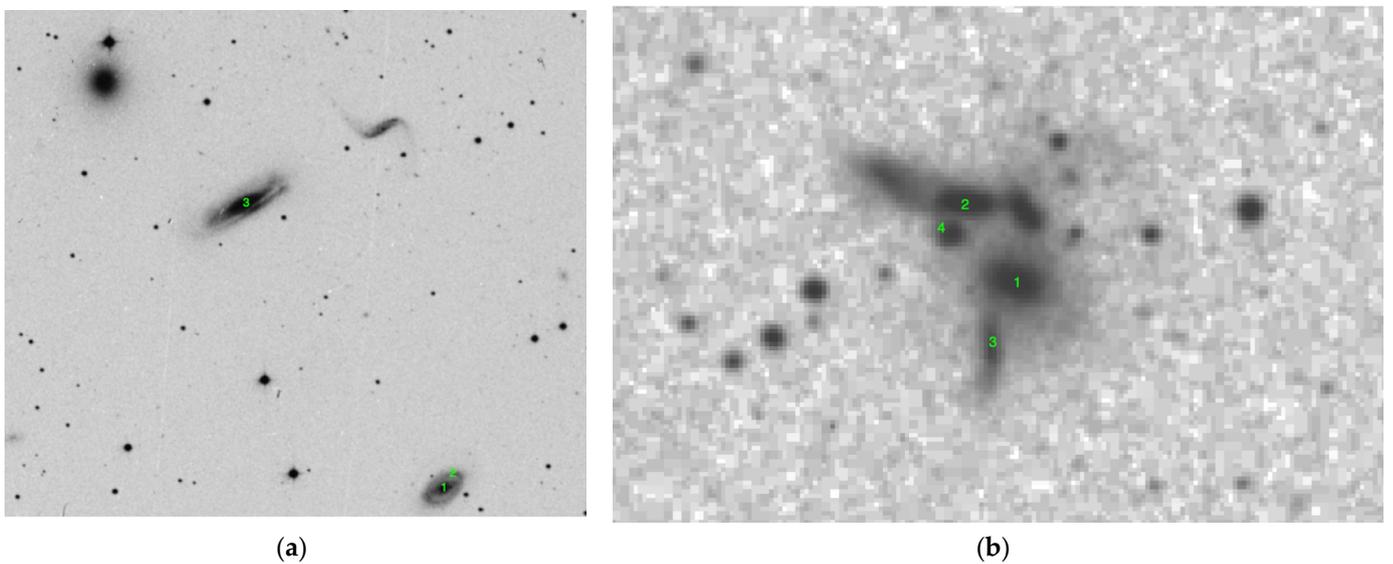

(**a**)      (**b**)

**Figure 3.** (**a**) XMM X-ray sources coincident with the galaxies on DSS image of HCG 44. (**b**) Chandra X-ray point sources on DSS image of HCG 79.

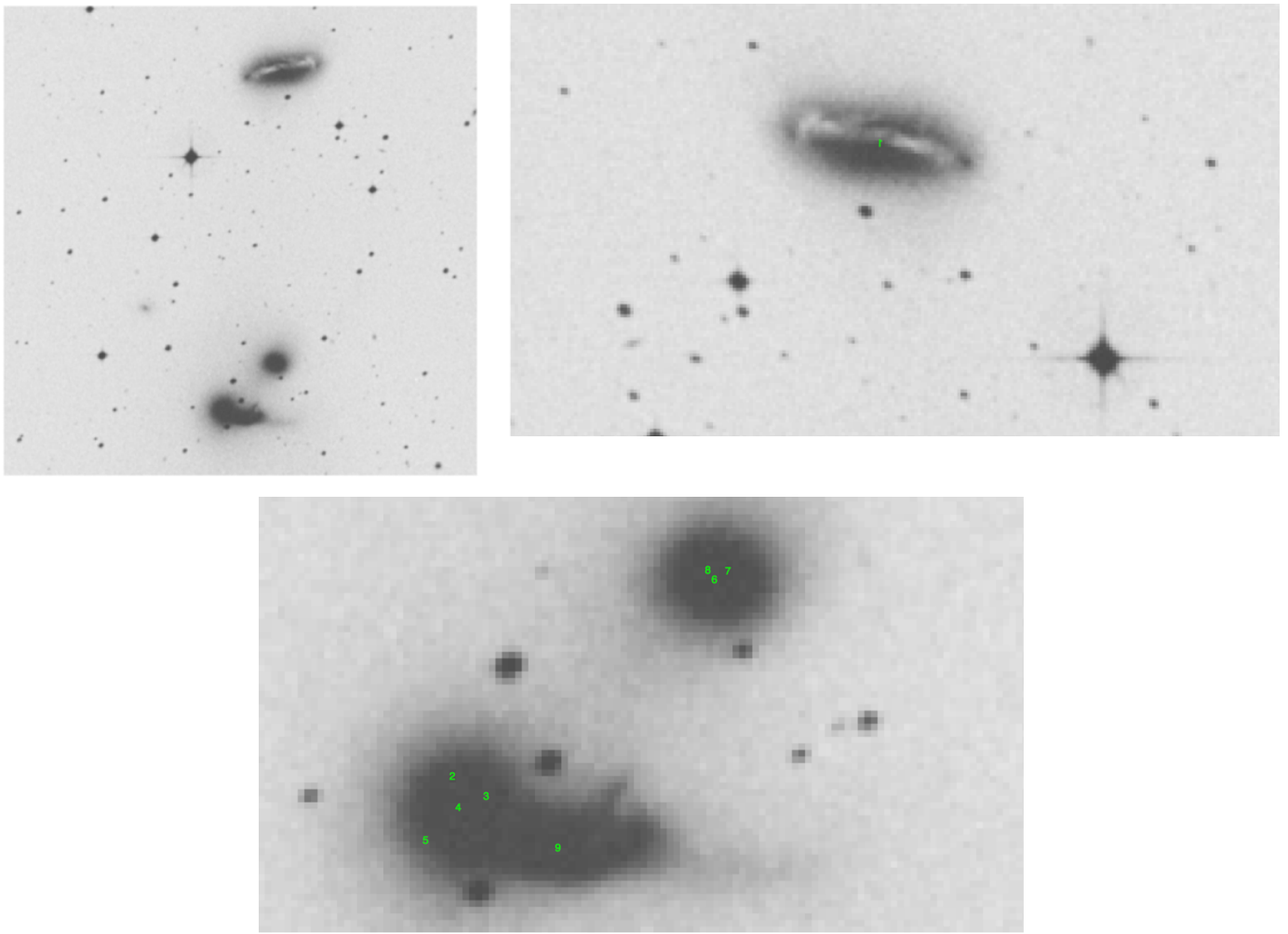

**Figure 4.** DSS image of HCG 90 (**top left**). X-ray point sources in NGC 7172 (**top right**). X-ray point sources in NGC 7176/NGC 7174 (**lower left—bottom image**) and NGC 7173 (**upper right—bottom image**).

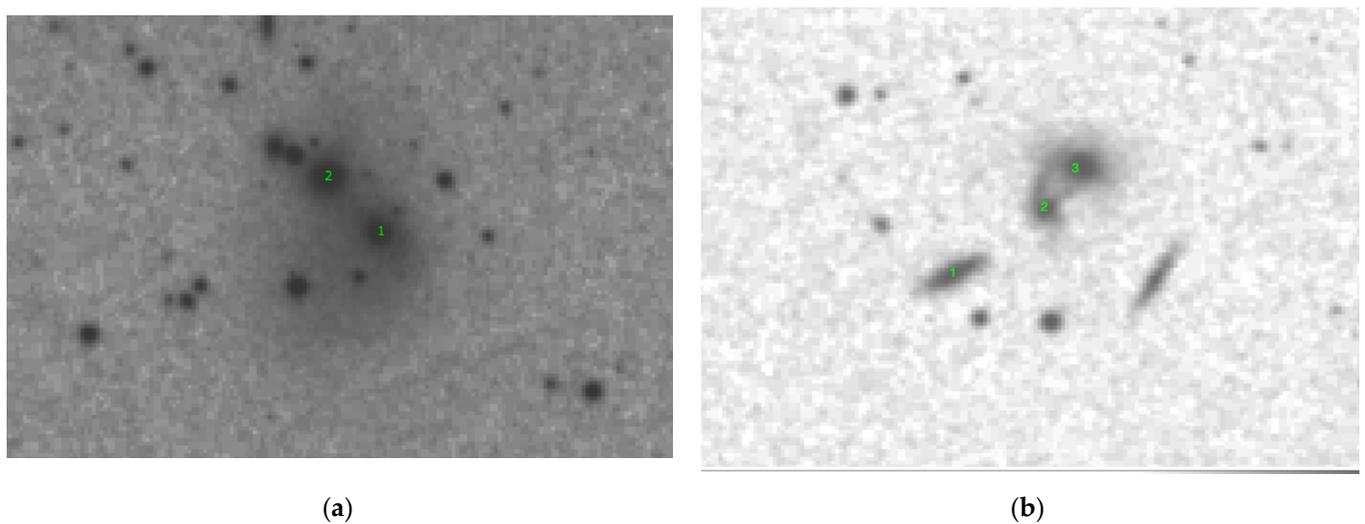

(**a**)                        (**b**)

**Figure 5.** (**a**) DSS image of HCG 94 with X-ray point sources. (**b**) DSS image of HCG 95 with X-ray point sources.

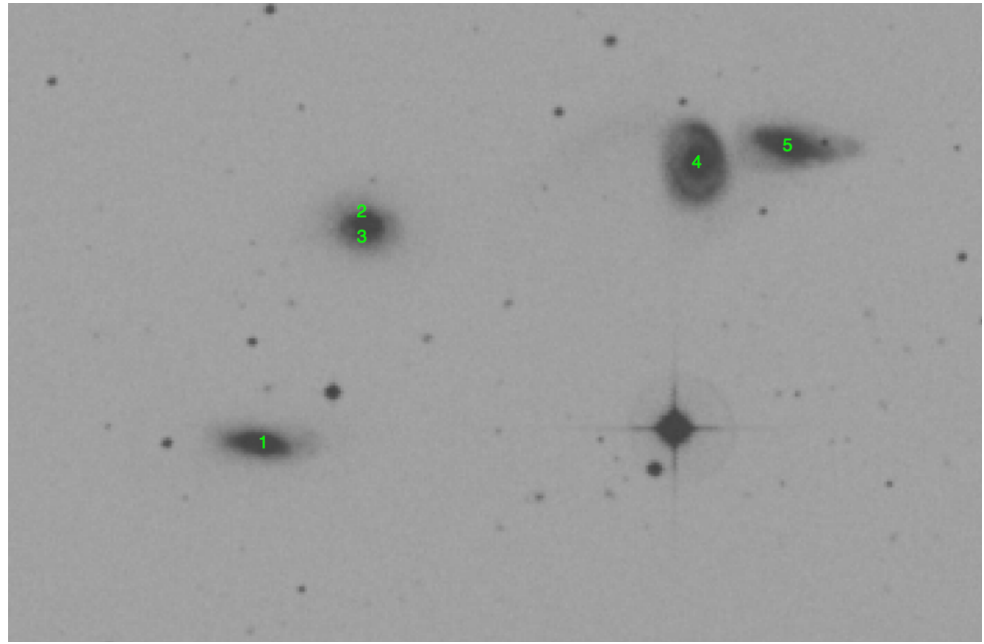

**Figure 6.** DSS image of HCG 16 with X-ray point sources.

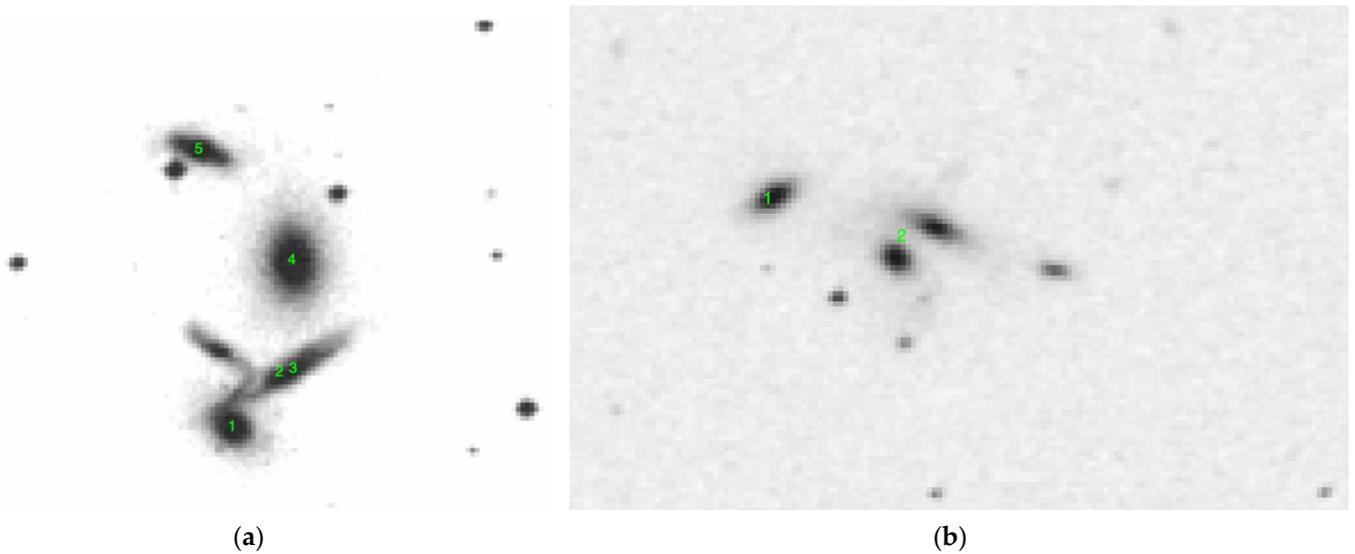

(**a**) (**b**)

**Figure 7.** (**a**) DSS image of HCG 40 with X-ray point sources. (**b**) DSS image of HCG 6 with X-ray point sources.

In this case, 59% of the sources are near the center of a galaxy while 41% appear to be out in the disk. From the data in the Tables 1 and 2, the X-ray luminosities are calculated using:

$$L_x = 4\pi D_H^2 (1+z)^2 (\text{Flux}),$$

where $D_H$ is the Hubble distance.

The X-ray luminosities of the point sources are shown in a histogram (Figure 8). An outlier X-ray point source of $4.14 \times 10^{42}$ erg s$^{-1}$ is in the core of the northern most galaxy in HCG 90 and not plotted. This source is likely a rather strong AGN. The luminosities are peaked around $10^{40}$ erg s$^{-1}$. The distribution around the peak is broad.

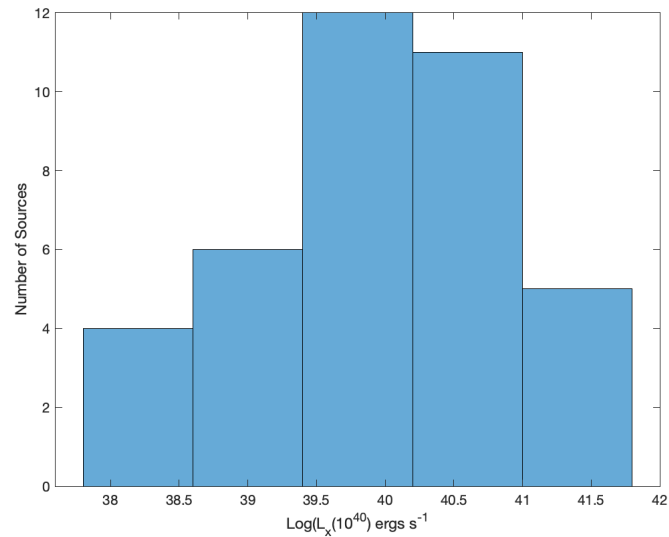

**Figure 8.** Histogram of X-ray luminosity for IGL sample. Most are $10^{39}$–$10^{41}$ erg s$^{-1}$.

*3.1 The NGC 5153 Group: A Close Pair Within a Group*

NGC 5153, an E1 pec elliptical galaxy, and NGC 5152, an SB(s)b type spiral, form a close pair, both in projection and with a velocity difference of 155 km s$^{-1}$. Additionally, the barred spiral, NGC 5150 may make this a group of three galaxies (see Figure 9). The basic X-ray source parameters are shown in Table 3.

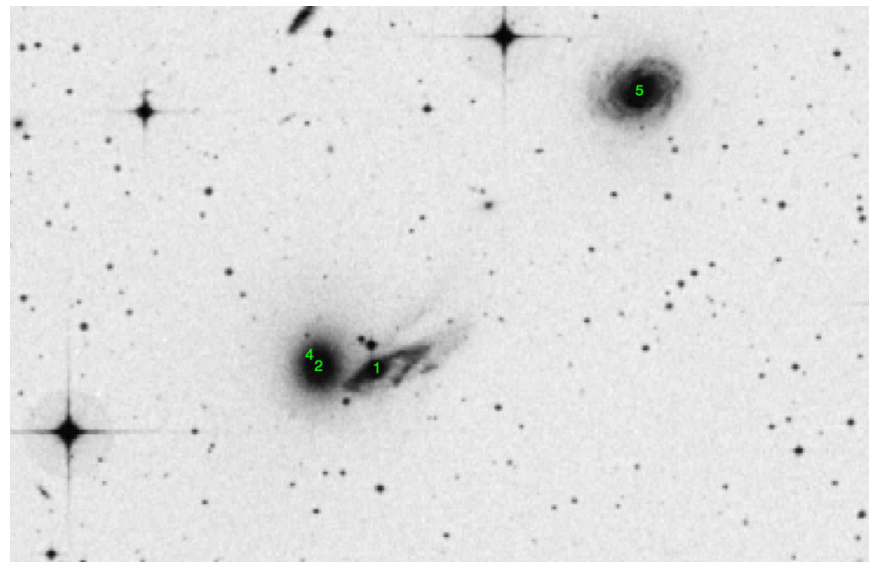

**Figure 9.** NGG 5153 Group DSS image with X-ray point sources.

**Table 3.** NGC 5153 Group Parameters.

| Galaxy | Counts | Error | Flux (×10$^{-14}$) | Lx (10$^{39}$) |
|---|---|---|---|---|
| NGC 5152-1 | 36.27 | 6.85 | 2.09 | 8.28 |
| NGC 5153-2 | 5.50 | 2.45 | 0.31 | 1.68 |
| NGC 5153-3 | 28.64 | 5.53 | 1.65 | 8.87 |
| NGC 5150-4 | 27.57 | 5.25 | 1.79 | 7.87 |

In an analysis of 8 compact groups [19], a majority of the galaxies show evidence of tidal interaction with a close neighbor. The mean velocity of a sample of mixed close pairs [20] is 199.2 km s$^{-1}$. The low relative velocity in close pairs, and in the 5152/5153 pair, in particular, means a longer interaction time. It implies substantial dissipation of orbital kinetic energy, indicative of an impending merger. In the general formation scheme of fossil group formation, the galaxy content goes from spiral dominated to spiral poor. In this scheme, the NGC 5153 group would be at an early stage as it is spiral dominated but clearly shows a future merger. An X-ray emitting intergalactic medium, as well as optical intergalactic light, develops in parallel with the changing galaxy content due to interactions. We expect to see a decrease in the level of star formation within groups that have IGL, because the IGL is a remnant of the interaction. The log of the total X-ray luminosity for the pair is 40.27, with a 14% error, significantly below the average for isolated mixed pairs 40.87 +/- 0.11. This may not be surprising as the close mixed pairs, due to their isolation, may be only one merger away from the formation of a massive elliptical while the NGC 5153 group may have an additional merger. The data set used for the pair study, from the ROSAT PSPC, was also more sensitive to extended emission from the hot IGM and sometimes unable to resolve point sources from extended emission, having 10s of arc sec resolution.

## 4. Discussion

We compare our sample of groups with intergalactic light (IGL) to a sample of 15 compact groups (CG) [21]. The selection criteria of the CG sample are similar to the IGL sample, with requirements on size and redshift. There are four compact groups common to both samples: HCG 16, 40, 79 and HCG 90. They have been removed from the CG sample for comparison. An outlier, a source in HCG 92 with luminosity $2.13 \times 10^{42}$ erg s$^{-1}$ that has been identified with an AGN [21] and has not been plotted or included in the statistical comparison. The distribution of X-ray luminosities for the CG sample is shown in Figure 10 (left panel). The distribution is highly peaked, similarly to IGL around $10^{40}$ erg s$^{-1}$, but narrow.

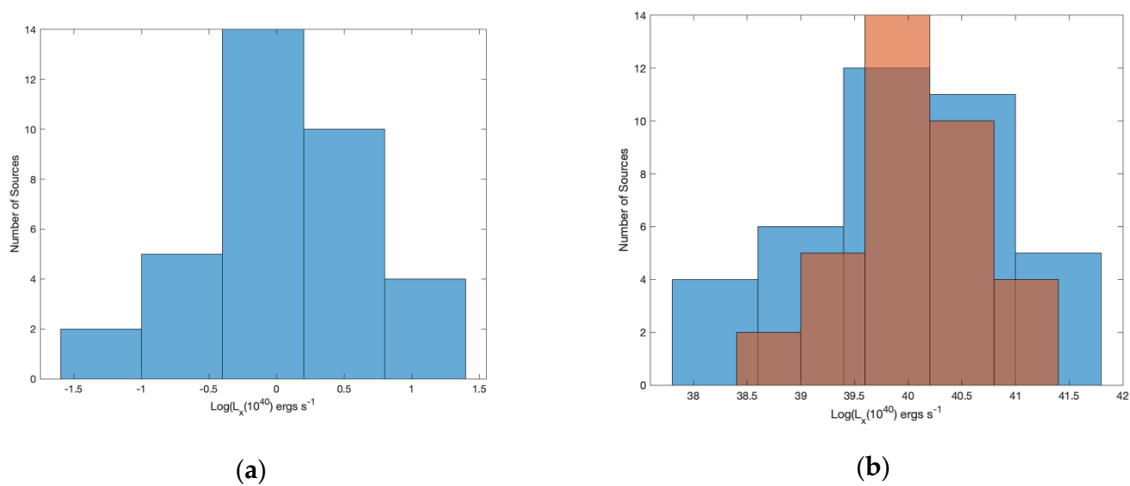

(a)      (b)

**Figure 10.** (**a**) Histogram of X-ray luminosities for compact groups *without* intergalactic light (CG). (**b**) The IGL luminosity histogram (in blue) and the CG luminosity histogram (in orange). The GC sample has narrow range and is more sharply peaked.

In Figure 10 (Right Panel), we show the two distributions, IGL and CG, superimposed. Visual inspection of the histograms shows that the IGL galaxies have more X-ray emission $< 3 \times 10^{39}$ erg s$^{-1}$ indicating galaxies with a lower SFR or perhaps low luminosity AGN sources in the core. The Kolmogorov-Smirnov Two Sample Test in *Matlab* is used to test the similarity of the IGL and CG samples. A probability of 0.26 is obtained; marginal evidence that the samples have a different X-ray point sources distribution. To further investigate this difference, we divide the IGL sample into two subsamples: core and disk. From visual inspection of Figures 2–7, sources are classified as core or disk (shown in Tables 1 and 2). There are 22 core sources and 16 disk sources. The idea is to separate core sources, which may be dominated by low luminosity AGN from disk sources, dominated by XRBs.

We first compare the distribution of IGL core X-ray sources with nuclear X-ray sources of nearby, normal galaxies. The sample of nuclear sources consists of 62 galaxies selected from the Spitzer Infrared Normal Galaxy Survey (SINGS) [22]. We will call the X-ray data for the 62 galaxies given in [22], XSINGS. The mean of the XSINGS is $8.00 \times 10^{39}$ compared to $5.68 \times 10^{40}$ for the IGL core sources, a factor of ~7 different. Clearly, in Figure 11, the histogram of IGL core sources is only co-spatial with the high end of XSINGS. The peak of the distributions differ by > 100. There are a couple contributing factors, both of which are a result of the difference in galaxy distances between the two samples. The group galaxies are further away with an average redshift of 0.0231, which gives a lookback time of 320 million years. Evolution of the super massive black hole driving the AGN might play a limited role in reducing the luminosity over this time so that the XSINGS, being nearby, are systematically lower. However, more importantly, the XSINGS sample is optically selected and has a number of upper limits (here treated as detections). Treating the upper limits as detections makes the KS test result more conservative compared to a statistical test that handles upper limits. This is because treating an upper limit as a detection raises the mean and makes the sample difference reported here less significant. The flux limit is about the same for both samples, $\sim 10^{-15}$ erg cm$^{-2}$ s$^{-1}$. As a result, since the IGL galaxies are significantly further away, they are biased toward the detection of higher X-ray luminosity. The low luminosity XSINGS sources, $< 10^{38}$, would be below the flux limit at the distances of the IGL galaxies and not detected. We conclude that evolution is likely not enough to account for the difference in X-ray luminosities but that low luminosity core sources are not detectable in the IGL sample, and this could account for the difference.

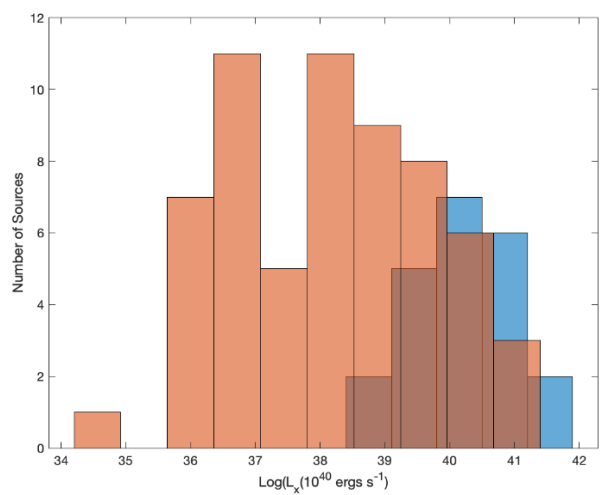

**Figure 11.** Comparison of nuclear X-ray sources in a SINGS sample of nearby galaxies (orange) to IGL core sources (blue). The samples are different with > 99.9% probability.

The IGL disk and core samples are shown in Figure 12. While the subsamples overlap, their distribution is somewhat different, with 89% probability. The core distribution is shifted to higher luminosity with a mean $5.68 \times 10^{40}$ compared to $1.85 \times 10^{40}$ for the disk. There are more high luminosity core sources and more low luminosity disk sources.

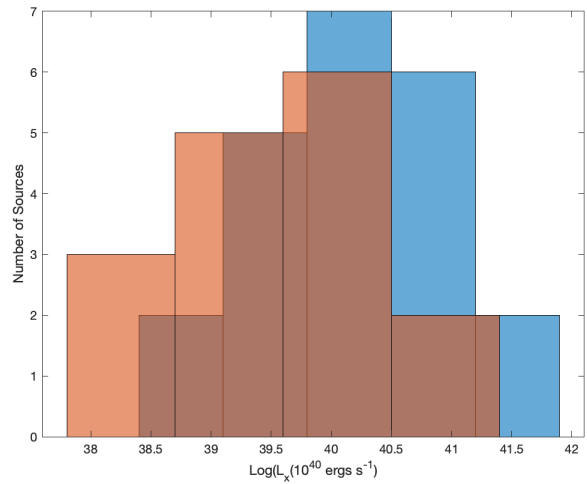

**Figure 12.** comparison of IGL core (blue) and disk (orange) sources.

In Figure 13, the IGL core and disk samples are compared separately to the CG sample. Here the differences between IGL and CG become more significant. The CG sample (without IGL) and the IGL disk sources differ with a probability of 94.4%. As can be seen in the right panel of Figure 13, the disk sources have a higher fraction of low luminosity sources relative to typical HCGs. As the disk sources are likely to be HMXRBs and indicative of SFR, this in interpreted as an indication that the groups with IGL, in general, have more galaxies that are post-starburst and in the quenching phase.

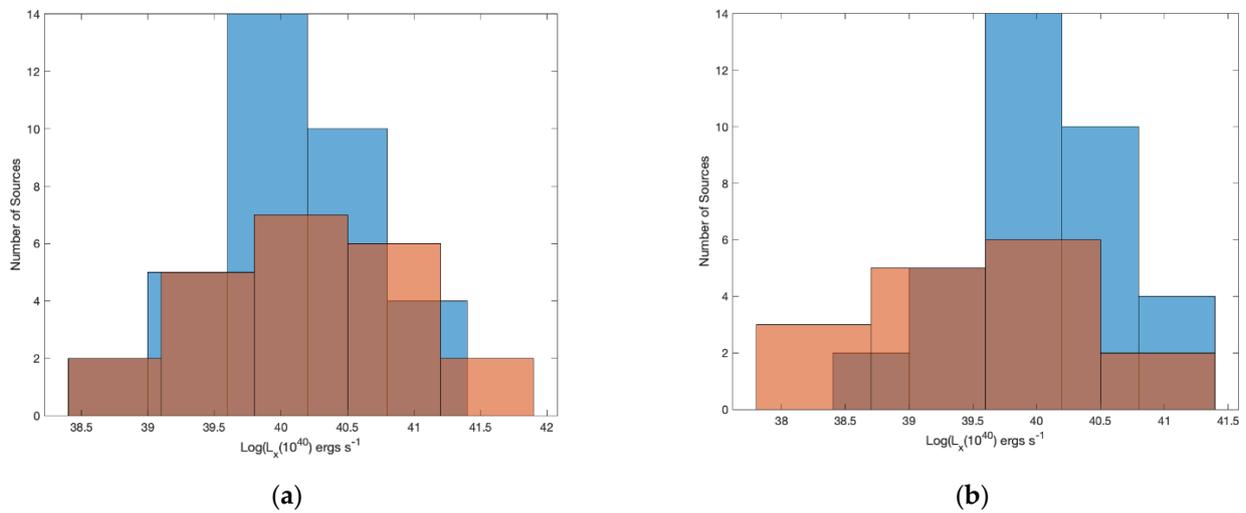

**Figure 13.** (**a**) Comparison of IGL core sources (orange) and the CG sample (blue). (**b**) comparison of IGL disk sources (orange) and the CG sample (blue).

The statistical test results for the various samples discussed are given in Table 4. The Kolmogorov-Smirnov (KS) test is used for testing whether or not two samples are drawn from the same parent sample. The percentages given in the table are (1-p)*100. Where p, the primary parameter returned from the KS test, is the probability that the samples are drawn from the same parent sample.

Table 4. Statistical Results.

| Sample | Mean X-ray Luminosity ($10^{40}$ erg s$^{-1}$) | Second Test Sample | Kolmogorov-Smirnov Test (Probability of Difference) |
| --- | --- | --- | --- |
| IGL | 4.06 | CG | 61% |
| IGL Core | 5.68 | - | - |
| IGL Disk | 1.85 | IGL Core | 89% |
| CG | 2.45 | IGL Disk | 94.4% |
| CG | - | IGL Core | 53% |
| SINGS Nuclear | 0.80 | IGL core | >99.9% |

Chandra observations of NGC 891 were used to obtain a galactic X-ray luminosity function for XRBs in a normal spiral galaxy [12]. They detected 26 binaries above a minimum flux of $10^{-15}$ erg cm$^{-2}$ s$^{-1}$. The slope of the luminosity function (XLF) is determined with 15% uncertainty. Compared with the XLF for XRBs in normal galaxies, the XLF in starburst galaxies is flatter. This is easily seen in the sample of various galaxies [12]. There appears to be two distinguishable groups: those with flat XLF slopes, ~0.5 and those with steeper LF slopes, ~1.0. The former group shows a lower 60μ/100μ flux ratio and includes: M82, NGC 4449, NGC 6503 and the Antennae–all well-known starbursts. The latter group includes normal spirals: NGC 1291, NGC 3184, IC 5332. There are several with intermediate XLF slope, ~ 0.7, they are also starbursts: NGC 253. These comparisons indicate that starburst galaxies have more high luminosity X-ray sources. The IGL sample, with a similar flux limit, is significantly more distant, and we mostly probe the HMXRB population and the ULXs. The luminosity of HMXRBs, up through the ULXs is well correlated with star formation rate [7]. So we are directly characterizing the SFR. Visual inspection of the normalized histogram in Figure 14 suggests that it has less power at $L_x > 10^{40}$ erg s$^{-1}$ compared with the CG galaxies. That is, typical compact group galaxies have a higher rate of star formation.

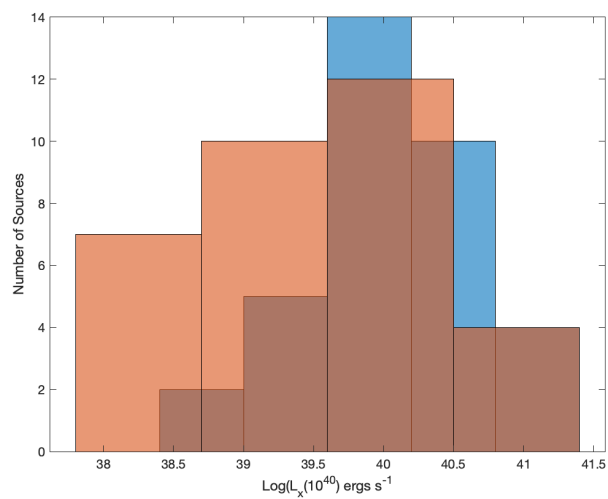

**Figure 14.** Approximately normalized histogram of IGL disk (orange) and CG (blue) sources.

There is a correlation between the fraction of elliptical galaxies and the amount of diffuse light in Hickson compact groups. Those with a large fraction of diffuse light are those with large fractions in number and luminosity of E/S0 galaxies. This indicates that the diffuse light is mainly created in dynamical processes during the formation of bright elliptical galaxies in major mergers [17]. Using N-body simulations, ~40% of the cluster's ICL is generated in the form of massive, dynamically cold streams. The production of streams requires strong tidal fields such as those due to close interaction between pairs of galaxies, and with merging of pairs of galaxies in clusters [23]. The simulations agree with observations, where 80 per cent of the stripped stars are placed in the ICL component and the other 20 per cent settles in the brightest cluster galaxy [24]. These processes must also be present in groups, and in particular, in the groups with IGL.

We found a greater number of low luminosity HMXRBs compared with typical compact groups. In the context of these simulations, our results imply a rapid quenching of the starburst phase. This is because all of the stripping processes that can account for the IGL have a phase of increased SFR and thus a bias toward ULXs. The absence of this bias in the IGL sample requires rapid quenching of the galaxies. A new population of recently quenched, blue ellipticals that are consistent with a recent merger origin, have been found in low mass groups [25]. In galaxy clusters, the ICL builds up rapidly around $z \sim 0.5$, and is the dominant evolutionary component of stellar mass in clusters from $z \sim 1$ [26]. [27] do not see a dependence between the presence of faint tidal features in compact groups and the brightness of their IGL. In addition, the IGL becomes brighter in the groups with a larger fraction of early-type galaxies. These results support the conclusion of this paper that accumulation of IGL is generally post-interaction and quiescence.

We also note that the gas ejected by galaxies that provides the diffuse optical intergalactic light, may be a significant process in the buildup of the warm, diffuse baryons associate with filaments [28], groups [29] and at large radius in clusters [30] as predicted by simulations [31,32].

## 5. Conclusions

We have characterized the X-ray point sources that are co-spatial with galaxies in compact groups known to have diffuse intergalactic light. The disk sources have X-ray luminosities typical of HMXRBs and ULXs. A comparison with typical compact groups shows a shift toward lower luminosity. Since HMXRB luminosity is proportional to the SFR, we infer that the galaxies in groups with IGL, in general, show a trend toward quiescence. This is consistent with simulations that show the presence of IGL being post-starburst, a phase that accompanies the gas removal mechanism accounting for IGL. Groups with IGL are a subset of compact groups in which merger toward a fossil group has begun.

Future work involves obtaining X-ray observations for two HCGs with IGL: HCG 35 and HCG 88 as well as redshifts for the two SCGs. Deep observations are required to characterize the nuclear AGN sources to determine if the profound statistical difference (IGL core sources are much stronger in the X-ray than normal galaxies) is just a selection effect. In addition, these observations will likely add to the completeness of the X-ray disk sample.

**Funding:** his research received no external funding.

**Data Availability Statement:** The data used in this article is available upon request.

**Conflicts of Interest:** The authors declare no conflict of interest.